\documentclass[aps,%
pre,%
 twocolumn,%
floatfix]{revtex4}
 \usepackage{epsfig}
\usepackage{here}

\begin{document}                % INITIALIZE - DONT CHANGE

\newcommand{\IGN}[1]{}

\newcommand{\Nullop}{{\bf 0}}
\newcommand{\Einsop}{{\bf 1}}

\title{Improved control of delayed measured systems}
\author{Jens Christian Claussen and Heinz Georg Schuster}
\affiliation{\mbox{Institut f\"ur Theoretische Physik der Universit\"at Kiel,
24098 Kiel, Germany}
\\
\mbox{\rm (Revised: July 29, 2004)}}
%\date{September 24, 1998}
%\date{September 24, 1998, revised \today}
%%\date{\today}
%\date{Revised: July 29, 2004}

\begin{abstract}                % DON'T CHANGE THIS LINE
In this paper we address the question how the control of 
delayed measured chaotic systems can be improved.
Both unmodified OGY control and difference control can be
successfully applied only for a certain range of Lyapunov numbers
depending on the delay time.
We show that this limitation can be overcome by at least two classes of methods,
by rhythmic control and by the memory methods 
linear predictive logging control (LPLC) and memory difference control (MDC).
\end{abstract}
\maketitle

%\narrowtext

\section{Introduction}
Delay is a generic problem in the control of chaotic systems.
The effective delay time $\tau$ in any feedback loop is the sum of at least 
three delay times, the duration of measurement, the time needed to compute 
the appropriate control amplitude, and the response time of the
system to the applied control. The latter effect appears especially when
the applied control additionally has to propagate through the system.
These response time may extend to one or more cycle lengths
\cite{mausbach95}.

For the formal situation of fixed point stabilization
in time-continuous control,
the issue of delay has been investigated
widely in control theory,
dating back at least to
the Smith predictor
\cite{smith57}.
This approach mimics the, yet unknown, actual
system state by a linear prediction based on the
last measurement.
Its time-discrete counterpart (LPLC) discussed below 
allows to place all eigenvalues of the associated
linear dynamics to zero, and always ensures stability.
The (time-continuous) Smith predictor
with its infinite-dimensional initial condition
had to be refined \cite{palmor80,hagglund96}, 
giving rise to the recently active fields of 
{\sl model predictive control}
% or {\sl receding horizon control}
\cite{mpc}.
For fixed point stabilization, an extension 
of permissible latency has been found
for a modified proportional-plus-derivative controller
\cite{sieber04}.

If one wants to stabilize the dynamics 
of a chaotic system onto an unstable periodic orbit,
one is in a special situation.
In principle, a proper engineering approach could be 
to use the concept of sliding mode control
\cite{slidingmode},
i.\ e.\ to use a co-moving coordinate system
and perform suitable control methods within it.
However, this requires the quite accurate knowledge of the whole trajectory 
(and 
%the local derivatives of the dynamical flow, to locate
 the direction of the stable manifold)
with the respective numerical or experimental costs.

Therefore direct approaches have been developed
by explicitely taking into account either
a Poincar\'e surface of section \cite{ogy90}
or the explicit periodic orbit length \cite{pyragas92}.
This field of {\sl controlling chaos}, 
or stabilization of chaotic systems, 
by small perturbations,
in system variables \cite{hubler}
or control parameters \cite{ogy90},
%has become
emerged to
 a widely discussed topic with applications
in a broad area from technical to biological systems.
Especially in fast systems 
\cite{socolar94,blakely}
or for slow drift in parameters \cite{claussen98a,mausbachpp99},
difference control methods have been successful,
namely the time-continuous Pyragas scheme
\cite{pyragas92},
ETDAS \cite{socolar94},
and time-discrete difference control
\cite{bielawski93a}.

Like for the control method itself, 
the discussion of the measurement delay
problem in chaos control 
has to take into account the special issues of the situation:
In classical control applications
one always tries to keep the control
loop latency as short as possible.
In chaotic systems however, 
one wants to control a fixed point 
of the Poincar\'e iteration
and thus has to wait until the next
crossing of the Poincar\'e surface of section,
where the system again is in vicinity of that fixed point.

The stability theory and the 
delay influence for time-continuous chaos control schemes
has been studied extensively
\cite{just97,just99pla,franc99,just99pre,hovel}, 
and an improvement of control by periodic modulation 
has been proposed in \cite{just03pre}.
For measurement delays that extend to a full
period, however no extension of the time-continuous 
Pyragas scheme is available.

In this paper we investigate time-discrete control schemes
and focus on the question what limitations occur if one
applies OGY control \cite{ogy90}
or difference feedback \cite{bielawski93a}
in the presence of time delay,
and what strategies can be used to overcome these limitations.
We show how the measurement delay
problem can be solved systematically
for OGY control and difference control
by rhythmic control and a memory method
and give constructive direct and elegant formulas 
for the deadbeat control in the time-discrete
Poincar\'e iteration.
While the predictive control method LPLC presented below 
for OGY control
has a direct correspondence to the Smith predictor
and thus can be reviewed as its somehow straightforward
implementation within the unstable subspace of the
Poincar\'e iteration,
this prediction approach does not guarantee a 
stable controller for difference control.
However, within a class of feedback schemes
linear in system parameters and system variable,
there is always a unique scheme where all eigenvalues
are zero, the MDC scheme presented below.
The method can be applied also for more than one positive Ljapunov
exponent, and shows, 
within validity of the linearization in vicinity of the orbit,
to be free of principal limitations in Ljapunov exponents or delay time.
For zero delay (but the inherent 1 period delay of MDC),
MDC has been demonstrated experimentally 
for a chaotic electronic circuit \cite{claussen98a}
and a thermionic plasma discharge diode \cite{mausbachpp99},
with excellent agreement, both of stability areas
and transient Ljapunov exponents, to the 
theory presented here.

The paper is organized as follows. 
After introducing the notation within a recall
of OGY control, 
we give a brief 
summary what limitations occur for unmodified
OGY control; 
details can be found in \cite{claussenjury}.
In Section \ref{secOGY}
we introduce different memory methods to improve control,
of which the LPLC approach appears to be superior as
it allows stabilization of arbitrary fixed points
for any given delay.
The stabilization of unknown fixed points is discussed in
Section 
%\ref{sec:diffkont} and
 \ref{sec:mdc}, 
where we present a memory method 
(MDC) that again allows stabilization
of arbitrary unstable fixed points.
As shown in Appendix~\ref{app1}, for all systems with only one
instable Lyapunov number, the iterated dynamics can be transformed
on an eigensystem which reduces to the one-dimensional case.
The explicit formulas for the case of higher-dimensional
subspaces are given in Appendices \ref{app_ogy} and \ref{app_diff}.

\section{Control of unstable periodic orbits}
\subsection{Ott, Grebogi, Yorke control}
The OGY method given by Ott, Grebogi and Yorke
\cite{ogy90}
stabilizes unstable fixed points (or unstable periodic orbits
utilizing a Poincar\'e surface of section) by feedback 
that is applied in vicinity of the fixed point $x^*$ 
of a discrete dynamics $x_{t+1}=f(x_t,r)$.

Thus, for a chaotic flow (or corresponding experiment)
one reduces the system dynamics
\begin{eqnarray}
\dot{\vec{x}} =\vec{F}(\vec{x},r) 
\end{eqnarray}
to the discrete dynamics between subsequent
Poincar\'e sections at $t_0, t_1, \ldots t_n$.
This description is fundamentally different from a stroboscopic sampling
as long as the system is not on a periodic orbit, where the
sequence of differences $(t_i-t_{i-1})$ would show a periodic structure.

If there is only one positive Ljapunov exponent,
we can proceed considering the motion in unstable
direction only (see Appendix \ref{app1}),
i.\ e.\ a one-dimensional iterated map.
(For 2 or more positive Ljapunov exponents
on can proceed in a similar fashion, see
App.\ \ref{app_ogy} and \ref{app_diff}.)

In OGY control, the control parameter $r_t$ is made time-dependent.
The amplitude of the feedback 
$r_t=r-r_0$
added to the control parameter $r_0$
is proportional by a constant $\varepsilon$ to the distance
$x-x^*$ from the fixed point, 
i.~e. $r=r_0 + \varepsilon (x_t-x^*)$, 
and the feedback gain
can be determined from a linearization around the fixed point,
which reads, if we neglect higher order terms,
\begin{eqnarray}
f(x_t,r_o+r_t)&=& f(x^{*},r_0)
+ (x_t-x^{*}) \cdot 
\left(\frac{\partial{} f}{\partial{} x}\right)_{x^{*},r_0}
\nonumber \\ & & 
\makebox[15em][l]{$
+ r_t \cdot
\left(\frac{\partial{} f}{\partial{} r}\right)_{x^{*},r_0}
$}
% + o(\ldots)
\nonumber 
\\
&=& 
\makebox[15em][l]{$
f(x^{*},r_0)+ \lambda (x_t-x^{*})+\mu r_t
$}
%+ o(\ldots)
\nonumber 
\\
&=& 
\makebox[15em][l]{$
f(x^{*},r_0)+ (\lambda + \mu \varepsilon)\cdot (x_t-x^{*})
$}
%+ o(\ldots)
\end{eqnarray}  
The second expression vanishes for $\varepsilon=-\lambda/\mu$,
that is, in linear approximation the system arrives 
at the fixed point at the next time step, $x_{t+1}=x^{*}$.
The uncontrolled system is assumed to be unstable in the fixed point,
% therefore we have the situation
i.\ e.\ $|\lambda|>1$.
The system with applied control is stable if the 
absolute value of the eigenvalues of the iterated map
is smaller than one,
\begin{eqnarray}
|x_{t+1}-x^{*}|=
|(\lambda + \mu \varepsilon)\cdot (x_{t}-x^{*})|
<|x_{t}-x^{*}|
\end{eqnarray}  
Therefore $\varepsilon$ has to be chosen 
between $(-1-\lambda)/\mu$ and $(+1-\lambda)/\mu$,
and this interval is of width $2/\mu$ and independent of $\lambda$,
i.e. fixed points with arbitrary $\lambda$ can be stabilized.
This property however does not survive for delayed measurement. 
One has to develop further control strategies to cover this class 
of systems.

%%%%%%%% new paragraph start

%\clearpage
\subsection{Delay matching in experimental situations}
Before discussing the time-discrete reduced dynamics 
in the Poincar\'e iteration, it should be clarified
how this relates to an experimental control situation.
On the first glance, the time-discrete viewpoint seems to
correspond only to a case where the delay 
(plus waiting time to the next Poincar\'e section)
 exactly matches the orbit length,
or a multiple of it. 
The generic experimental 
situation however comes up with a non-matching delay.  
Application of all control methods discussed here 
requires to introduce an additional delay,
usually by waiting for the next Poincar\'e crossing,
so that measurement and control are applied 
without phase shift at the same position of the 
orbit. 
In this case the next Poincar\'e crossing
position $x_{t+1}$ is a function of 
the values of $x$ and $r$ at a finite number of
previous Poincar\'e crossings only, 
i.~e. it does not depend on intermediate positions.
Therefore the (a priori infinite-dimensional)
delay system reduces to a finite-dimensional
iterated map.

If the delay (plus the time of the
waiting mechanism to the next Poincar\'e crossing)
is not matching the orbit length,
the control schemes may perform less efficient.
Even for larger deviations from the orbit, the
time between the Poincar\'e crossings
will vary only marginally, thus a
control amplitude should be available in time.
In practical situations therefore the
delay should not exceed the 
orbit length minus the variance of the
orbit length that appears in the respective
system and control setup.

In a formal sense, 
the Poincar\'e approach 
ensures robustness with respect to 
uncertainties in the orbit length,
as it always ensures a synchronized reset
of both trajectories and control.
Between the Poincar\'e crossings the control
parameter is constant, the system is independent of
everything {\sl in advance of} the
last Poincar\'e crossing.
It is solely determined by the 
differential equation (or experimental dynamics).
Thus the next crossing position is
a well defined iterated function of the 
previous one.

This is quite in contrast to the situation of a
delay-differential equation (as in Pyragas control),
which has an infinite-dimensional initial condition
it `never gets rid of'.
One may proceed to stability analysis via
Floquet theory 
\cite{hale}
as investgated for
% various 
continuous \cite{just97}
and Poincar\'e-based \cite{claussenthesis,claussenfloquet} control schemes.
Though a Poincar\'e crossing detection may 
be applied as well, the position will depend not only 
on the last crossing, but also on 
all values of the system variable within a time horizon 
defined by the maximum of the delay length
and the (maximal) time difference between two Poincar\'e crossings
(due to the non-stroboscopic character).
Thus the Poincar\'e iteration would be a
function between two infinite-dynamical spaces. 
Apart from further mathematical subtleties,
for a delay differential equation with fixed delay
the major advantage of a Poincar\'e map,
reducing the system dynamics
to a low-dimensional system, thus completely breaks down.

For all control schemes discussed within this paper,
howvever the additional dimensionality is not
a continuous horizon of states, but merely a finite
set of values that were measured at the 
previous Poincar\'e crossings.

%\clearpage

\IGN{
In any case, a time-continuous (Floquet) stability analysis 
of the respective differential equations
including the delays (or iteratively retarded values)
is appropriate, and stability borders have to be
extracted from transcendent characteristic equations.
E.\ g.\ one can analyze 
the interesting case of the influence of the
control impulse length \cite{claussenfloquet}.
For small delays there is a 
continuous, at first linear, decay in the
stability range \cite{claussenthesis,claussenfloquet}
similar to the results 
of Just et al.\
\cite{just????}
for the Pyragas type control schemes.
A well known related problem 
 the delay mismatch
effect for 
 Pyragas control
\cite{just??}.
An investigation of nonideal experimental
Poincar\'e realizations might be interesting,
though highly system-dependent.
The reported experiments
\cite{claussen98a,mausbachpp99} of MDC 
however agreed well with the time-discrete picture,
same as hundreds of applications of the 
also Poincar\'e-based OGY scheme.
}%IGN

%%%%%%%% new paragraph end

\section{OGY control: Delayed case, rhythmic and simple memory control}
\label{secOGY}

\subsection{Unmodified control of delayed measured maps}
To illustrate the problem, we recall the simple case
where no modification of the OGY scheme is taken
into account \cite{claussenjury}.
For $\tau$ time steps delay, unmodified proportional feedback
is applied:
\begin{eqnarray}
r_t = \varepsilon (x_{t-\tau}-x^*).
\end{eqnarray}  
Without loss of generality, we can choose $r_t=0$
if no control is applied, 
and $x^*=0$ in the remainder. 
Using the time-delayed coordinates
$(x_t, x_{t-1}, x_{t-2},
\ldots x_{t-\tau})^{\mbox{\scriptsize\rm{}T}}$:
\begin{eqnarray}
\!\!
\left(
\!\!
\begin{array}{c}
x_{t+1}\\ \vdots\\\\\\\\\\\vdots\\x_{t-\tau+1}
\end{array}
\!\!
\right)
\!\!
= 
\!\!
\left(
\!\!
\begin{array}{ccccccc}
\lambda&0&\cdots&&\cdots&0&\varepsilon\mu\\
1&0&&&&&0\\
0&1&\ddots&&&&\vdots\\
\vdots&&\ddots&&&&\\  
&&&&\ddots&&\\
\vdots&&&&\ddots&0&\vdots\\
0&\cdots&&\cdots&0&1&0
\end{array}
\!\!
\right)
\!\!
\left(
\!\!
\begin{array}{c}
x_{t}\\ \vdots\\\\\\\\\\\vdots\\x_{t-\tau}
\!\!
\end{array}
\!\!
\right)
\label{eq:matrixogyunmodified}
\end{eqnarray}
To ensure stability, one has to fulfill  $|\alpha_i|<1$
for all eigenvalues $\alpha_i$. 
The stability area is investigated in \cite{claussenjury},
yielding a $\tau$-dependent maximal Ljapunov number of
\begin{eqnarray}
\lambda_{\rm max} = 1 + \frac{1}{\tau}.
\label{lmaxvontau}
\label{eq:lambdamaxogy}
\end{eqnarray}
For $\lambda\geq\lambda_{\rm max}(\tau)$
no control is possible \cite{claussenjury}.
%\begin{eqnarray}
%\Lambda_{\rm max} \cdot \tau \leq 1.  
%\label{eq:controllability}
%\end{eqnarray}

\IGN{
In general, if the system is measured delayed by $\tau$ steps,
\begin{eqnarray}
r_t = \varepsilon x_{t-\tau},
\end{eqnarray}  
we write the dynamics in delayed coordinates
$(x_t, x_{t-1}, x_{t-2},
\ldots x_{t-\tau})^{\mbox{\scriptsize\rm{}T}}$:
\begin{eqnarray}
\!\!
\left(
\!\!
\begin{array}{c}
x_{t+1}\\ \vdots\\\\\\\\\\\vdots\\x_{t-\tau+1}
\end{array}
\!\!
\right)
\!\!
= 
\!\!
\left(
\!\!
\begin{array}{ccccccc}
\lambda&0&\cdots&&\cdots&0&\varepsilon\mu\\
1&0&&&&&0\\
0&1&\ddots&&&&\vdots\\
\vdots&&\ddots&&&&\\  
&&&&\ddots&&\\
\vdots&&&&\ddots&0&\vdots\\
0&\cdots&&\cdots&0&1&0
\end{array}
\!\!
\right)
\!\!
\left(
\!\!
\begin{array}{c}
x_{t}\\ \vdots\\\\\\\\\\\vdots\\x_{t-\tau}
\!\!
\end{array}
\!\!
\right)
\label{eq:matrixogyunmodified}
\end{eqnarray}

The characteristic polynomial is given by
\begin{eqnarray}
0=(\alpha-\lambda)\alpha^{\tau}-\varepsilon\mu,
\end{eqnarray}
or, rescaled by 
$\tilde{\alpha} := \alpha/\lambda$
and $\tilde{\varepsilon} = \varepsilon\mu/\lambda^{\tau+1}$
we have
\begin{eqnarray}
0=(\tilde{\alpha}-1)\tilde{\alpha}^{\tau}
-\tilde{\varepsilon}.
\end{eqnarray}

}%IGN

%\typeout{Fig 1 goes here!}

\IGN{
\begin{figure}[htbp]
\begin{center}
\epsfig{file=kurven4.eps, angle=270,width=3in}
\end{center}
\caption{
Control intervals for several time delays $\tau=0\ldots{}5$:
The plots show the maximal absolute value of the 
rescaled eigenvalues $|\tilde{\alpha}|$
as a function of the rescaled control gain $\tilde{\varepsilon}$.
For $\tau=0$, an appropriate value of $\tilde{\varepsilon}=-1$
ensures all eigenvalues to be zero (OGY without delay).
For $\tau>0$ the maximal $|\lambda|$ that can be controlled control 
is given by the minimum of the corresponding plot by
$|\tilde{\alpha}_{\rm{}min}|=1/|\lambda|$.
\label{fig:kurven4}}
\end{figure}
}%IGN

\IGN{

In Fig.~\ref{fig:kurven4} we have plotted the maximum of the
absolute value of the eigenvalues, for $\tau=0,1,\ldots,5$.
In rescaled coordinates $\tilde{\alpha}=1/\lambda$ corresponds to
a control interval $\tilde{\varepsilon}_{\pm}(\tau,\lambda)$.
For
\begin{eqnarray}
\lambda_{\rm max} = 1 + \frac{1}{\tau}
\label{lmaxvontau}
\label{eq:lambdamaxogy}
\end{eqnarray}
this control interval vanishes, and for $\lambda\geq\lambda_{\rm max}(\tau)$
no control is possible.

For the conditional Lyapunov exponent 
 (or the absolute value of the Floquet multiplier of the
 motion around that orbit determined by the
 underlying differential equation)
$\Lambda:=\ln |\lambda|$
of the orbit,
we have by $\ln x \leq (x-1)$ the inequality
\begin{eqnarray}
\Lambda_{\rm max} \cdot \tau \leq 1.  
\label{eq:controllability}
\end{eqnarray}
Therefore, delay time and Lyapunov exponent limit each other 
if the system is to be controlled.
This controllability relation is consistent with the loss of knowledge in the system
by exponential separation of trajectories.

For small $\tau$ one can derive the borders of the
stability area in a staightforward manner
with help of the Jury criterion \cite{jury}.
This is done in \cite{claussenjury}.
The stability area is shown in Fig.~\ref{fig:tau1234quadrant}.

}%IGN

\typeout{Fig.2 goes here!}

\IGN{
\begin{figure}[htbp]
\begin{center}
\epsfig{file=tau1234quadrant.eps, angle=270,width=3in}
\end{center}
\caption[quadrant]{
Stability areas for delayed, but unmodified
 OGY control for $\tau=1, 2, 3, 4$, combined.
Only for $|\lambda|>1$ control is necessary (dashed line),
and the stability area extends to
$\lambda_{\rm{}max}= 2, 3/2, 4/3, 5/4$.
%%%(ordinata: $0\leq|\lambda|\leq +2$ and
%%%abscissa:
%%%$0\leq|\mu\varepsilon|\leq +1$).
\label{fig:tau1234quadrant}}
\end{figure}
}

\subsection{Rhythmic OGY control}
\label{sec:mem}
As pointed out for difference control 
in the case $\tau=0$ in 
\cite{bielawski93a,claussenthesis,claussenphyscon,schusterstemmler}, 
one can eliminate the additional degrees of freedom 
caused by the delay term. 
One can 
restrict himself to 
apply control rhythmically only every $\tau+1$ timesteps
($\tau+2$ for difference control), and then leave the system uncontrolled
for the remaining timesteps.
Then $\varepsilon=\varepsilon(t)$ appears to be time-dependent with
\begin{eqnarray}
\varepsilon ({t {\rm ~  mod ~ } \tau}) = (\varepsilon_0,0,\ldots,0)
\end{eqnarray}
and, after $(\tau+1)$ iterations of (\ref{eq:matrixogyunmodified}), 
we again have a matrix as in (\ref{eq:matrixogyunmodified}),
but with $\lambda^{\tau+1}$ instead of $\lambda$.
Equivalently, we can write 
\begin{eqnarray}
x_{t+(\tau+1)}=\lambda^{\tau+1}x_t+\varepsilon_0\mu x_t.
\end{eqnarray}
What we have done here, is: controlling the $(\tau+1)$-fold
iterate of the original system.
This appears to be formally elegant, but leads to
practically uncontrollable high effective
Lyapunov numbers $\lambda^{\tau+1}$ for both large $\lambda$
and large~$\tau$.

Even if the rhythmic control method is of striking simplicity,
it remains unsatisfying that control is kept quiet,
or inactive, for $\tau$ time steps. 
Even if the state of the system $x$ is known delayed by
$\tau$, one knows (in principle) the values of $x_t$ for
$t<\tau$, and one could (in principle) store the values
$\delta{}r_{t-\tau}\ldots{}\delta{}r_{t}$ of the control
amplitudes applied to the system. 
This can be done, depending on the timescale, by analog 
or digital delay lines, or by storing the values in a computer
or signal processor (there are some 
intermediate
frequency ranges where an
experimental setup is difficult).

Both methods, rhythmic control and simple feedback 
control in every time step, have their disadvantages.

For rhythmic control it is necessary to use rather large 
control amplitudes, in average $\lambda^\tau / \tau$, 
and noise sums up to an amplitude increased by factor $\sqrt{\tau}$.

For simple feedback control the dimension of the system
is increased and the maximal controllable Lyapunov number
is bounded by (\ref{lmaxvontau}).

One might wonder if there are control strategies 
that avoid these limitations.
This has necessarily to be done by applying control in each time step,
but with using knowledge what control has been applied 
between the last measured time step $t-\tau$ and~$t$.
This concept can be implemented in at least two ways,
by storing previous values of $x_t$ or previous values of 
$\delta r_t$ (LPLC).

\subsection{Control using memory for previous states}
The first memory method extends the single delay line to
several delay lines, using one control gain coefficient for each:
\begin{eqnarray}
r_t &=& \varepsilon_1  x_{t-1}
 + \varepsilon_2  x_{t-2} 
 + \ldots 
+  \varepsilon_{n+1}  x_{t-n-1} 
\end{eqnarray}

For $n$ steps memory (and one step delay) the control matrix is 
\begin{eqnarray}  
\!\!
\left(
\!\!
\begin{array}{c}
x_{t+1}\\ \vdots\\\\\\\\\\\vdots\\x_{t-n}
\end{array}
\!\!
\right)
\!\!
= 
\!\!
\left(
\!\!
\begin{array}{ccccccc}
\lambda& \varepsilon_1
&\cdots&& &\varepsilon_{n}& \varepsilon_{n+1}\\
1&0&&&&&0\\
0&1&\ddots&&&&\vdots\\
\vdots&&\ddots&&&&\\  
&&&&\ddots&&\\
\vdots&&&&\ddots&0&\vdots\\
0&\cdots&&\cdots&0&1&0
\end{array}
\!\!
\right)
\!\!
\left(
\!\!
\begin{array}{c}
x_{t}\\ \vdots\\\\\\\\\\\vdots\\x_{t-n-1}
\end{array}
\!\!
\right)
\end{eqnarray}
with the
characteristic
polynomial
\begin{eqnarray}
(\alpha-\lambda)\alpha^{n+1} + \sum_{i=1}^{n} \varepsilon_i \alpha^{n-i}
\end{eqnarray}
We can choose $\alpha_1 = \alpha_2 = \ldots \alpha_{n+2} = -\lambda/(n+2)$
and evaluate optimal values for all $\varepsilon_i$ 
by comparing with the coefficients of the product
$\prod_{i=1}^{n+2} (\alpha-\alpha_i)$.
This method allows control up to $\lambda_{\rm{}max}=2+n$,
therefore arbitrary $\lambda$ can be controlled if a memory
length of $n>\lambda-2$ and the optimal coefficents $\varepsilon_i$
are used.

For more than one step delay, one has the situation
$\varepsilon_1=0,\ldots,\varepsilon_{\tau-1}=0$.
This prohibits the 'trivial pole placement' given above,
(choosing all $\alpha_i$ to the same value) 
and therefore reduces the maximal controllable $\lambda$
and no general scheme for optimal selection of the $\varepsilon_i$ applies.
One can alternatively use the LPLC method described below,
%%%%%%%% new paragraph start
which provides an optimal control scheme.
One could wonder why to consider the previous state memory scheme at all
when it does not allow to make all eigenvalues zero in {\sl any} case.
First, the case of up to one orbit delay and moderately small
$\lambda$ already covers many low-period orbits.
Second, there may be experimental setups where the feedback of 
previous states through additional delay elements and 
an analog circuit is experimentally more feasible
than feedback of controlled amplitudes.

In summary, all three methods discussed in this section have principal
restrictions, but may be of advantage in special situations especially
when a simple setup is required.
%%%%%%%% new paragraph end

\subsection{Linear predictive logging control (LPLC)} \label{sec:lplc}
If it is possible to store the previously applied control amplitudes
$r_t, r_{t-1},\ldots$, then one can predict the actual state $x_t$ of 
the system using the linear approximation around the fixed point.
That is, from the last measured value $x_{t-\tau}$ 
and the control amplitudes 
we compute estimated values iteratively by
\begin{eqnarray}
y_{t-i+1}  =  \lambda x_{t-i} + \mu  r_{t-i} 
\end{eqnarray}
leading to a {\it predicted} value $y_t$ of the actual system state.
Then the original OGY formula can be applied, i.~e.
$r_t= -\lambda/\mu y_t$.
In this method the gain parameters are again linear in $x_{t-\tau}$ and
all $\{ r_{t'}\}$ with $t-\tau \leq t' \leq t$, and the optimal
gain parameters can be expressed in terms of $\lambda$ and $\mu$.

In contrast to the memory method presented in the previous subsection,
the LPLC method directs the system (in linear approximation) in one
time step onto the fixed point.
However, when this control algorithm is switched on, one has no control
applied between $t-\tau$ and $t-1$, so the trajectory has to be fairly
near to the orbit (in an interval with a length 
of order $\delta/\lambda^\tau$, where $\delta$ is the 
interval halfwidth where control is switched on).
Therefore the time one has to wait until the control can be successfully
activated is of order $\lambda^{\tau-1}$ larger than in the case of
undelayed control.

The LPLC method can also be derived as a general linear feedback
in the last measured system state and all applied control
amplitudes since the system was measured; choosing the feedback
gain parameters so that the linearized system has all eigenvalues 
zero.
The linear ansatz
\begin{eqnarray}
r_{t} = 
\varepsilon \cdot
x_{t-\tau-i} 
+ \eta_1  r_{t-1} + \ldots  \eta_{\tau}  r_{t-\tau} 
\end{eqnarray}
leads to the dynamics in
combined delayed coordinates
\\
$(x_t, x_t-1, \ldots ,x_{t-\tau}, r_{t-1},  \ldots , r_{t-\tau})$
\footnotesize
\begin{eqnarray}
\!\!
\left(
\!\!
\begin{array}{c}
x_{t+1}\\ 
x_t\\ \vdots \\ \vdots \\  \vdots \\ x_{t-\tau+1}\\
 r_{t} \\ \vdots \\ \vdots \\  \vdots \\ r_{t-\tau+1}
\end{array}
\!\!
\right)
\!\!
= 
\!\!
\left(
\!\!
\begin{array}{ccccccccccc}
\lambda&0&\cdots&\cdots&0&\varepsilon&  \eta_1&\eta_2&\cdots&\cdots&\eta_{\tau} \\
 & 0 & & & & & & & & & \\
 & 1 & \ddots & & & & & & & & \\
 & & \ddots&\ddots & & & & & & & \\  
 & & & \ddots&\ddots & & & & & & \\  
 & & & & 1& 0& & & & & \\
0 &0&\cdots&\cdots&0&\varepsilon& \eta_1&\eta_2&\cdots&\cdots&\eta_{\tau} \\
 & & & & & &1 &0 & & & \\  
 & & & & & & &\ddots &\ddots & & \\  
 & & & & & & & & \ddots&\ddots & \\  
 & & & & & & & & &1 &0 
\end{array}
\!\!
\right)
\!\!
\left(
\!\!
\begin{array}{c}
x_t\\ 
x_{t-1}\\ \vdots \\ \vdots \\  \vdots \\ x_{t-\tau}\\
 r_{t-1} \\ \vdots \\ \vdots \\  \vdots \\ r_{t-\tau}
\end{array}
\!\!
\right)
\nonumber
\end{eqnarray}
%\begin{eqnarray}   \end{eqnarray}
\normalsize
\noindent
giving the characteristic polynomial
\begin{eqnarray}
0 &=& -\alpha^\tau
(
\alpha^{\tau+1}
+\alpha^{\tau} (-\lambda - \eta_1)
\nonumber\\
\nonumber
& &
\mbox{~~~~~}
+\alpha^{\tau-1} (\lambda \cdot \eta_1 - \eta_2)
+\alpha^{\tau-2} (\lambda \cdot \eta_2 - \eta_3) 
\\
&& \mbox{~~~~~} \ldots
+\alpha^{1} (\lambda \cdot \eta_{\tau-1} - \eta_{\tau}) 
+ (\lambda \cdot \eta_{\tau} -\varepsilon)   ).
\end{eqnarray}
All eigenvalues can be set to zero using
$\varepsilon=-\lambda^{\tau+1}$
and $\eta_i=-\lambda^{i}$.
The general formulas even for more than one positive
Lyapunov exponent or 
multiparameter control are given in Appendix~\ref{app_ogy}.

\subsection{Nonlinear predictive logging control}
One can also consider a 
nonlinear predictive logging control (NLPLC)
strategy as the 
straightforward extension to the LPLC method for nonlinear
prediction. If the system has a delay of several time steps, 
the interval where control is achieved becomes too small.
However, if it is possible to extract the first nonlinearities from the 
time series, prediction (and control) can be fundamentally improved. 
In NLPLC the behaviour of the system is predicted each time step by
\begin{eqnarray}  
x_{t+1}  =  \lambda x_t +\frac{\lambda_2}{2} x_t^2 
 + \mu r_t + \frac{\mu_2}{2} r_t^2 
\nonumber 
+ \nu x_t r_t
+ o(x_t^3,\ldots) 
\end{eqnarray}
with using the applied control amplitudes $\{r_t\}$ for each time step.
This equation has to be solved for $r_t$ using $x_{t+1} \stackrel{!}{=}0$.

A similar nonlinear prediction method has been described by 
Petrov and Showalter \cite{showalter96}.
They approximate the $x_{t+1}(x_t,r_t)$ surface 
directly from the time series and use it to direct the system
to any desired point.
Both Taylor approximation or Petrov and Showalter method 
can be used here iteratively, provided one knows the delay length.

Both approaches could be regarded as a nonlinear method of model
predictive control \cite{mpc},
applied to the Poincar\'e iteration dynamics.

From a practical point of view, 
it has to be mentioned that one has to know the fixed point $x^*$
more accurate than in the linear case.
Otherwise one experiences a smaller range of stability
and additionally a permanent nonvanishing control
amplitude will remain.
This may be of disadvantage especially
if the fixed point drifts in time (e.g. by other external parameters
such as temperature) or if the time series used to determine the 
parameters is too short.

\section{Stabilization of unknown fixed points}
%\label{sec:diffkont}
\label{sec:mdc}
As all methods mentioned above require the knowledge of the 
position of the fixed point, one may wish to stabilize 
purely by feeding back differences of the system variable 
at different times.
Without delay, difference feedback can be used successfully for 
$\varepsilon\mu=-\lambda/3$, and eigenvalues of modulus smaller 
than unity of the matrix
{\small
$
\left(
\begin{array}{cc}
\lambda +\varepsilon\mu & - \varepsilon\mu  \\ 1 & 0
\end{array}
\right)
$
\normalsize}
are obtained only for $-3 < \lambda < +1 $, so this method stabilizes 
only for oscillatory repulsive fixed points with $-3 < \lambda < -1$
\cite{bielawski93a,schusterstemmler}.

Due to the inherent one period delay of MDC,
the $\tau$ period delay case of
MDC corresponds,
in terms of the number of degrees of freedom,
to the  $\tau+1$ period delay
case of LPLC.

\subsection{Unmodified difference control of iterated delayed measured maps}
In the presence of $\tau$ steps delay 
the linearized dynamics of
a simple difference feedback
$r_t-r_0 = \varepsilon (x_{t-\tau}-x_{t-\tau-1})$
is given by
\begin{eqnarray}
\!\!
\left(
\!\!
\begin{array}{c}
x_{t+1}\\ \vdots\\\\\\\\\\\vdots\\x_{t-\tau}
\end{array}
\!\!
\right)
\!\!
= 
\!\!
\left(
\!\!
\begin{array}{ccccccc}
\lambda&0&\cdots&&0&\varepsilon\mu&-\varepsilon\mu\\
1&0&&&&&0\\
0&1&\ddots&&&&\vdots\\
\vdots&&\ddots&&&&\\
&&&&\ddots&&\\
\vdots&&&&\ddots&0&\vdots\\
0&\cdots&&\cdots&0&1&0
\end{array}
\!\!
\right)
\!\!
\left(
\!\!
\begin{array}{c}
x_{t}\\ \vdots\\\\\\\\\\\vdots\\x_{t-\tau-1}
\end{array}
\!\!
\right)
\!\!
\end{eqnarray}
in delayed coordinates
$\vec{y}_{t}=(x_t,x_{t-1},\ldots x_{t-\tau-1})$.
%and the characteristic polynomial is given by
%\begin{eqnarray}
%0=(\alpha-\lambda)\alpha^{\tau}+ (1-\alpha) \varepsilon\mu.
%\end{eqnarray}
%
As we have to use $x_{t-\tau-1}$  in addition to 
$x_{t-\tau}$, the system is of dimension $\tau+2$,
and the lower bound of Lyapunov numbers that can be controlled
are found to be \cite{claussenjury}
\begin{eqnarray}
\lambda_{\rm{}inf} = - \frac{3+2\tau}{1+2\tau}.
\label{eq:lambdamaxdiff}
\end{eqnarray}
The controllable range is smaller than for 
unmodified OGY control, and is restricted
to oscillatory repulsive fixed points with 
$\lambda_{\rm{}inf} < \lambda \leq -1$.
A striking observation is that inserting $\tau+\frac{1}{2}$ for $\tau$
in eq.~(\ref{eq:lambdamaxogy}) leads exactly to the expression in
eq.~(\ref{eq:lambdamaxdiff}) which reflects the fact that the difference
feedback can be interpreted as a discretized first time derivative, taken at
time $t-\tau-\frac{1}{2}$.
For details and stability area see \cite{claussenjury}. 

\typeout{Fig.3 goes here!}

\IGN{
\begin{figure}[htbp]
\begin{center}
\epsfig{file=jurydiff_tau_0123.eps, angle=270 ,width=3in}
\end{center}
\caption[Jury difference tau 01 combined]{
Stability of difference feedback 
without improvements for $\tau=0,1,2,3$:
Stability borders are derived by the Jury criterion.
For $|\lambda|<1$ the system ist stable without control, and
the range of $\lambda$ that can be stabilized decreases 
with increasing $\tau$.
%%%(ordinata: $\lambda$, abscissa: $\mu\varepsilon\cdot(-1)^{\tau}$)
\label{fig:jurydiff01}}
\end{figure}
}

\subsection{Rhythmic difference control}
\begin{figure}[htbp]
\begin{center}
\epsfig{file=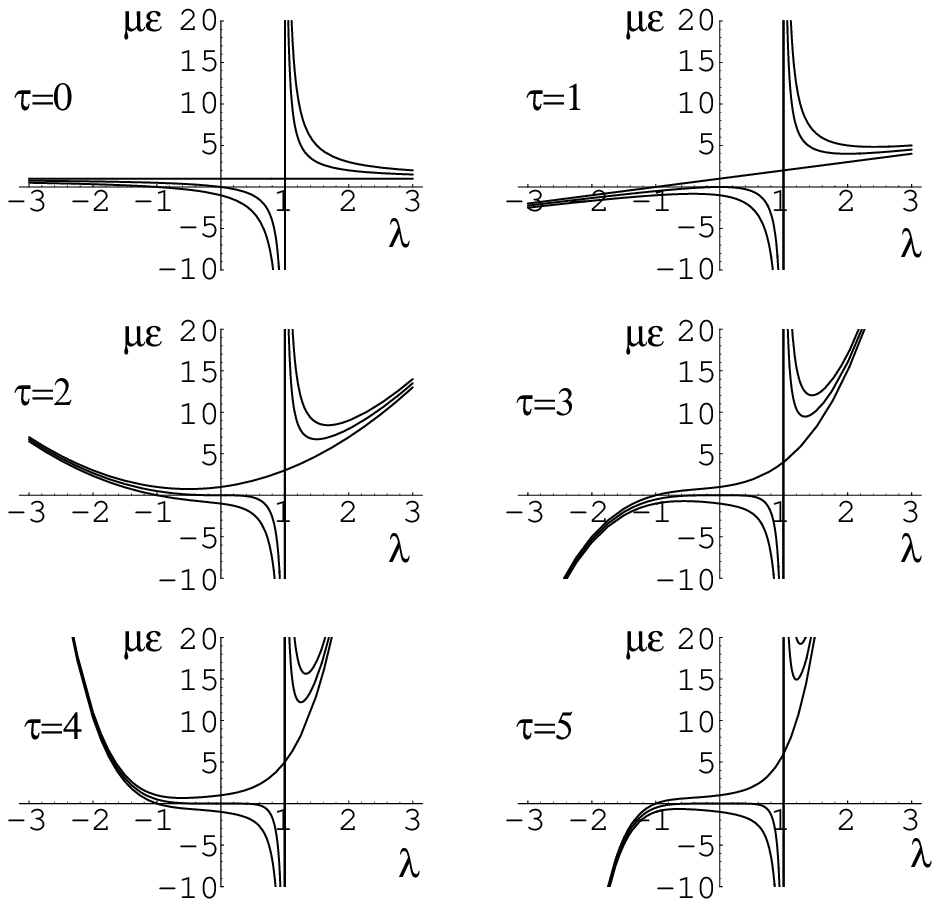, angle=0, width=82mm}
\end{center}  
\IGN{
\begin{center}
\epsfig{file=diff_tau0.eps, angle=270, width=1.4in}
\epsfig{file=diff_tau1.eps, angle=270, width=1.4in}
\end{center}  
\begin{center}
\epsfig{file=diff_tau2.eps, angle=270, width=1.4in}
\epsfig{file=diff_tau3.eps, angle=270, width=1.4in}
\end{center}
\begin{center}
\epsfig{file=diff_tau4.eps, angle=270, width=1.4in}
\epsfig{file=diff_tau5.eps, angle=270, width=1.4in} 
\end{center}
}%IGN
\caption[periodic difference tau 0 1 2 3 4 ]{
Periodic difference feedback for $\tau=0,1,2,3,4,5$:
Maximal, optimal and minimal value of $\mu\varepsilon$
for given $\lambda$
to obtain stabilization by control applied every $\tau+2$ time
steps. 
%%%  (ordinata: $\mu\varepsilon$, abscissa: $\lambda$)
\label{fig:peri_difftau012345}}
\end{figure}

To enlarge the range of controllable $\lambda$, one
again has the possibility 
to reduce the dimension of the
control process in linear approximation
to one by applying control every $\tau+2$ time steps.
\begin{eqnarray}
x_{t+1}
&=& 
\lambda x_t +
\mu \varepsilon (x_{t-\tau}-x_{t-\tau-1})
\label{eq:peridiff_opt}
\\
&=&
(\lambda^{\tau+1} +\mu \varepsilon \lambda - \mu \varepsilon)
x_{t-\tau-1}
\nonumber
\end{eqnarray}
and the goal $x_{t+1} \stackrel{!}{=} 0$ can be fulfilled by
\begin{eqnarray}
\mu \varepsilon =
- \frac{\lambda^{\tau+1}}{1-\lambda}
\end{eqnarray}
One has to choose $\mu \varepsilon$ between 
$\mu \varepsilon_{\pm}= - \frac{\lambda^{\tau+1} \pm 1}{1-\lambda} $
to achieve control as shown in Fig.~\ref{fig:peri_difftau012345}.
The case $\tau=0$ has already been discussed in
\cite{bielawski93a,claussenthesis,claussenphyscon,schusterstemmler}.
With rhythmic control,
there is no range limit for $\lambda$, and even fixed points
with positive $\lambda$ can be stabilized by this method.

When using differences for periodic feedback, one
still has the problem that the control gain increases by $\lambda^\tau$,
and noise sums up for $\tau+1$ time steps before the next control
signal is applied.
Additionally, now there is a singularity for $\lambda=+1$ in the ``optimal''
control gain given by (\ref{eq:peridiff_opt}).
This concerns fixed points where differences $x_t-x_{t-1}$ 
when escaping from the fixed point are naturally
small due to a $\lambda$ near to $+1$.

Here one has to decide between using a large control gain 
(but magnifying noise and finite precision effects)
or using a small control gain of order 
$\mu\varepsilon_{-}(\lambda=+1) =\tau+1$ 
(but having larger eigenvalues and therefore slow convergence).

%\IGN{
%\tiny
%\begin{verbatim}
%Plot[{-(x^1-1)/(1-x),-(x^1)/(1-x),-(x^1+1)/(1-x)},{x,-3,3}, PlotRange -> {-10,20}] 
%Plot[{-(x^2-1)/(1-x),-(x^2)/(1-x),-(x^2+1)/(1-x)},{x,-3,3}, PlotRange -> {-10,20}]
%Plot[{-(x^3-1)/(1-x),-(x^3)/(1-x),-(x^3+1)/(1-x)},{x,-3,3}, PlotRange -> {-10,20}]
%Plot[{-(x^4-1)/(1-x),-(x^4)/(1-x),-(x^4+1)/(1-x)},{x,-3,3}, PlotRange -> {-10,20}]
%Plot[{-(x^5-1)/(1-x),-(x^5)/(1-x),-(x^5+1)/(1-x)},{x,-3,3}, PlotRange -> {-10,20}]
%Plot[{-(x^6-1)/(1-x),-(x^6)/(1-x),-(x^6+1)/(1-x)},{x,-3,3}, PlotRange -> {-10,20}]
%p7=Plot[{-(x^1-1)/(1-x),-(x^1)/(1-x),-(x^1+1)/(1-x)},{x,-3,3}, PlotRange -> {-10,20}, AxesLabels->{me,Text[FontForm[l,{Symbol,12}]]}] 
%Show[%1,Epilog->{Text[FontForm[l,{Symbol,12}]],{-.5,20}}]
%\end{verbatim}
%\normalsize
%} %footnote

\IGN{
p0=Plot[{-(x^1-1)/(1-x),-(x^1)/(1-x),-(x^1+1)/(1-x)},{x,-3,3}, PlotRange -> {-10,20}] 
p1=Plot[{-(x^2-1)/(1-x),-(x^2)/(1-x),-(x^2+1)/(1-x)},{x,-3,3}, PlotRange -> {-10,20}]
p2=Plot[{-(x^3-1)/(1-x),-(x^3)/(1-x),-(x^3+1)/(1-x)},{x,-3,3}, PlotRange -> {-10,20}]
p3=Plot[{-(x^4-1)/(1-x),-(x^4)/(1-x),-(x^4+1)/(1-x)},{x,-3,3}, PlotRange -> {-10,20}]
p4=Plot[{-(x^5-1)/(1-x),-(x^5)/(1-x),-(x^5+1)/(1-x)},{x,-3,3}, PlotRange -> {-10,20}]
p5=Plot[{-(x^6-1)/(1-x),-(x^6)/(1-x),-(x^6+1)/(1-x)},{x,-3,3}, PlotRange -> {-10,20}]
pall=Show[GraphicsArray[{{p0,p1},{p2,p3},{p4,p5}}]]
Display["Untitled2.eps",pall,"EPS"]
}

\typeout{Fig. 4 goes here!}

%\IGN{
%}%IGN

Two other strategies that have been discussed by Socolar and Gauthier
\cite{socolar98} are discretized versions of time-continuous methods.
Control between $\lambda= -(3+R)/(1-R)$  and  $\lambda= -1$ is possible with
discrete-ETDAS ($R<1$)
%\begin{eqnarray}
$
r_t=\varepsilon \sum_{k=0}^\infty R^k (x_{t-k}-x_{t-k-1})
$
%\end{eqnarray}
and control between $\lambda= -(N+1)$ and  $\lambda= -1$ 
is acheived with discrete-NTDAS (let $N$ be a positive integer)
%\begin{eqnarray}
$
r_t=\varepsilon \left(x_t-\frac{1}{N}\sum_{k=0}^N x_{t-k}\right).
$
%\end{eqnarray}
Both methods can be considered to be of advantage even in 
time-discrete control in the Poincar\'e section, e.g. if the
number of adjustable parameters has to be kept small.
Whereas these methods are mainly applied in time-continuous 
control, especially in analogue or optical experiments,
for time-discrete control the MDC strategy described below
allows to overcome the limitations in the Lyapunov number.

\subsection{Memory difference control (MDC)} 
%\label{sec:mdc}
%
One may wish to generalize the linear predictive feedback
to difference feedback. 
In contrary to the LPLC case, 
the reconstruction of the state $x_{t-\tau}$ from 
differences $x_{t-\tau-i} - x_{t-\tau-i-1}$ 
and applied control amplitudes $r_{t-j}$ is no longer unique.
As a consequence, there are infinitely many ways to compute an estimate
for the present state of the system,
but only a subset of these leads to a controller design ensuring convergence
to the fixed point.

Under these there exists an optimal every-step control for difference
feedback with minimal eigenvalues and in this sense optimal stability.
This memory difference control (MDC) 
method has been demonstrated in an electronic experiment
\cite{claussen98a}
and a plasma diode \cite{mausbachpp99}.

To derive the feedback rule, 
we directly make the linear ansatz
\begin{eqnarray}
r_{t} = 
\varepsilon \cdot
(x_{t-\tau-i} - x_{t-\tau-i-1})
+ \eta_1  r_{t-1} + \ldots  \eta_{\tau}  r_{t-\tau} 
\end{eqnarray}
leading to the dynamics in
combined delayed coordinates
\begin{eqnarray}
\nonumber
(x_t+1, x_t, \ldots ,x_{t-\tau+1}, r_{t-1},  \ldots ,
r_{t-\tau+1})^{\mbox{\scriptsize\rm T}}
~~~~~~~~
\\
~~~~~~~
=
{\bf M}
\cdot
(x_t, x_t-1, \ldots ,x_{t-\tau}, r_{t-1},  \ldots , r_{t-\tau}
)^{\mbox{\scriptsize\rm T}}
\end{eqnarray}
with
\footnotesize
\begin{eqnarray} 
{\bf M} = 
\!\!
\left(
\!\!
\begin{array}{ccccccccccc}
\lambda&0&\cdots&0 &\varepsilon & -\varepsilon&  
\eta_1& \eta_2-\eta_1&\cdots&\cdots&\eta_{\tau}-\eta_{\tau+1} \\
1 & 0 & & & & & & & & & \\
 & 1 & \ddots & & & & & & & & \\
 & & \ddots&\ddots & & & & & & & \\  
 & & & \ddots&\ddots & & & & & & \\  
 & & & & 1& 0& & & & & \\
0&0&\cdots&0 &\varepsilon & -\varepsilon&  
\eta_1& \eta_2-\eta_1&\cdots&\cdots&\eta_{\tau} -\eta_{\tau+1} \\
 & & & & & &1 &0 & & & \\  
 & & & & & & &\ddots &\ddots & & \\  
 & & & & & & & & \ddots&\ddots & \\  
 & & & & & & & & &1 &0 
\end{array}
\!\!
\right)
\nonumber
\end{eqnarray}
\normalsize

\noindent
giving the characteristic polynomial
\begin{eqnarray}
0 &=& -\alpha^\tau
(
\alpha^{\tau+1}
+\alpha^{\tau} (-\lambda - \eta_1)
\nonumber
\\ &&
+\alpha^{\tau-1} (\lambda \cdot \eta_1 - \eta_2)
+\alpha^{\tau-2} (\lambda \cdot \eta_2 - \eta_3) 
\nonumber
\\ &&
 \ldots
+\alpha^{2} (\lambda \cdot \eta_{\tau-2} - \eta_{\tau-1}) 
\nonumber
\\ &&
+\alpha^{1} (\lambda \cdot \eta_{\tau-1} - \eta_{\tau}- \varepsilon) 
+ (\lambda \cdot \eta_{\tau} +\varepsilon)   ).
\end{eqnarray}
All eigenvalues can be set to zero using
$\varepsilon=-\lambda^{\tau+1}/(l-1)$
$\eta_{\tau}= + \lambda^{\tau}/(l-1)$
and $\eta_i=-\lambda^{i}$ for $1\leq i \leq \tau-1$.
\\
The general formulas even for more than one positive 
Lyapunov exponent or 
multiparameter control are given in Appendix~\ref{app_diff}.

%\clearpage
\section{Conclusions}
We have presented methods to improve 
Poincar\'e-section based chaos control for 
delayed measurement. For both classes of algorithms,
OGY control and difference control, delay 
affects control, and improved control strategies
have to be applied.
Improved strategies contain one of the following 
principle ideas: Rhythmic control, control with memory
for previous states, or control with memory
for previously applied control amplitudes.

In special cases the unmodified control, 
previous state memory control, or rhythmic control
methods could be considered, especially
when experimental conditions restrict the possibilities 
of designing the control strategy.

In general, the proposed LPLC and MDC strategies allow a so-called
deadbeat control with all eigenvalues zero; and they are
in this sense optimal control methods.
All parameters needed for controller design can be 
calculated from linearization parameters that can be
fitted directly from experimental data.
This approach has also been sucessfully 
applied in an electronic
\cite{claussen98a}
and plasma 
\cite{mausbachpp99}
experiment.

%\clearpage
\appendix
\section{Transformation on the eigensystem}
\label{app1}
Here we derive how for one unstable dimension
the stabilization problem reduces to the one-dimensional case.
Using a covariant basis from
right eigenvectors $\vec{e}_{\rm u}$
and $\vec{e}_{\rm s}$ 
to eigenvectors $\lambda_{\rm u}$ und
$\lambda_{\rm s}$
and the corresponding contravariant left
eigenvectors $\vec{f}^{\rm u}$ and $\vec{f}^{\rm s}$
of matrix $L$.
we can transform the linearized dynamics
\begin{eqnarray}
\vec{x}_{t+1} = L \vec{x}_t + M \vec{r}_t
\end{eqnarray}
with help of 
$\Einsop=\vec{e}_{\rm u} \cdot\cdot \vec{f}^{\rm u}
+ \vec{e}_{\rm s} \cdot\cdot \vec{f}^{\rm s}$
and $L= \lambda_{\rm u} \vec{e}_{\rm u} \cdot\cdot \vec{f}^{\rm u}
+\lambda_{\rm s}  \vec{e}_{\rm s} \cdot\cdot \vec{f}^{\rm s}$.
We define as coordinates in the eigensystem
$x^{\rm u}_t:=\vec{f}^{\rm u} \vec{x}_t$
and
$x^{\rm s}_t:=\vec{f}^{\rm s} \vec{x}_t$
giving
\begin{eqnarray}
x^{\rm u}_{t+1} &=& \lambda_{\rm u} x^{\rm u}_t + r_t \mu_{\rm u}
\nonumber\\
x^{\rm s}_{t+1} &=& \lambda_{\rm s} x^{\rm s}_t + r_t \mu_{\rm s}
\end{eqnarray}
We consider a general ansatz that the control signal 
is linear in 
 $x_{t-j}$ 
and $r_{t-i}$,
\begin{eqnarray}
 r_t = \sum_{j=0}^{\infty} \vec{K}_j \cdot \vec{x}_{t-j}
+  \sum_{i=0}^{\infty} \eta_i \cdot r_{t-i}
\end{eqnarray}
Here $r_t = \vec{K}\cdot \vec{x}_{t}$ is OGY control.
For $r_t$ follows
\begin{eqnarray}
 r_t &=& \sum_{j=0}^{\infty} \vec{K}_j 
(\vec{e}_{\rm u} \cdot\cdot \vec{f}^{\rm u}
+ \vec{e}_{\rm s} \cdot\cdot \vec{f}^{\rm s})\vec{x}_{t-j}
+  \sum_{i=0}^{\infty} \eta_i \cdot r_{t-i} 
\nonumber\\&=&
\sum_{j=0}^{\infty} (\vec{K}_j\cdot \vec{e}_{\rm u}) 
x^{\rm u}_{t-j}
+ \sum_{j=0}^{\infty} (\vec{K}_j\cdot \vec{e}_{\rm s})
x^{\rm s}_{t-j}
+  \sum_{i=0}^{\infty} \eta_i \cdot r_{t-i} 
\nonumber\\&=&
\sum_{j=0}^{\infty} K_j^{\rm u}
x^{\rm u}_{t-j} 
+ \sum_{j=0}^{\infty} K_j^{\rm s}
x^{\rm s}_{t-j}
+  \sum_{i=0}^{\infty} \eta_i \cdot r_{t-i} 
\end{eqnarray}
If control in stable direction is chosen zero, i.e.
 $\forall_{j}K^{\rm s}_{j}=0$,
the dynamics decouples in two two systems in stable
direction (that is unaffected, but was stable before)
and in unstable direction,
\begin{eqnarray}
x^{\rm u}_{t+1} =
 \lambda_{\rm u} x^{\rm u}_t +  \mu_{\rm u}
(\sum_{j=0}^{\infty} K_j^{\rm u} x^{\rm u}_{t-j}
+  \sum_{i=0}^{\infty} \eta_i \cdot r_{t-i}).
\end{eqnarray}
%
% weglassen?
% For OGY control the dynamics in unstable direction is simply
% \begin{eqnarray}
% x^{\rm u}_{t+1} =  \lambda_{\rm u} x^{\rm u}_t + \mu_{\rm u}
% (\vec{K}\cdot \vec{x}_{t}).
% \end{eqnarray}
%
This equation is only one-dimensional, so it is sufficient
to investigate the one-dimensional control problem
if there is only one instable direction.
The generalization to higher-dimensional unstable subspaces 
 is straightforward.

\section{Stabilization of delayed maps}
\label{app_ogy}
We consider the motion around an unstable fixed point $\vec{x}^*$ of 
the map 
\begin{eqnarray}
{\vec{x}_{t+1}} =\vec{F}(\vec{x}_t,\vec{r}_t)
\end{eqnarray}
where $\vec{r}_t$ and $\vec{x}_t$ have the same dimension of the phase space of
the system (although it is desirable to achieve control with a minor number
of control parameters).
The linear time evoution around the instable fixed point 
(we choose $\vec{x}^*=\vec{0}$ and $\vec{r}=\vec{0}$ in the fixed point)
is given by Jacobians $D_x=:L$ and $D_r=:M$
\begin{eqnarray}
\vec{x}_{t+1}= L \vec{x}_t + M \vec{r}_t
\end{eqnarray}
Delayed measurement means that the applied control $\vec{r}_t$ 
can only be computed from an older value $\vec{x}_{t-\tau}$.
Additionally one can store the control amplitudes
$\vec{r}_1{\ldots}\vec{r}_{t-\tau}$ applied in the meantime to the system
and use the linearization to predict the present position $\vec{x}$
(Linear predictive logging control, LPLC \cite{claussenthesis}).

Thus we have the general feedback ansatz
\begin{eqnarray}
\vec{r}_t = K \vec{x}_{t-\tau} + \sum_{j=1}^{\tau} N_j \vec{r}_{t-j}.
\end{eqnarray}
Under the condition that $\det{}M\neq{}0$, the feedback can be chosen to
\begin{eqnarray}
K &=& - M^{-1} L^{\tau+1} \\
\forall_{1\leq i \leq \tau} \;\;\;\;
N_i &=& - M^{-1} L^{i} M 
\end{eqnarray}
Iterating
$\vec{x}_t = L^\tau x_{t-\tau} + \sum_{j=1}^{\tau} 
L^{j-1} M \vec{r}_{t-j}$, 
one has 
\begin{eqnarray}
\vec{x}_{t+1}&=& L \vec{x}_t + M \vec{r}_t
 =  L^{\tau+1}  x_{t-\tau}
-M M^{-1}  L^{\tau+1}  x_{t-\tau}
\nonumber \\
& & 
 + \sum_{j=1}^{\tau} L^{j-1} M \vec{r}_{t-j}
-\sum_{j=1}^{\tau} M M^{-1}  L^j M \vec{r}_{t-j} 
 =  \vec{0}.
\nonumber
\end{eqnarray}

\section{Delayed measured maps -- Stabilizing unknown fixed points}
\label{app_diff}
Now we show that even fixed points whose exact position 
is not given (only an approximative value is needed to determine
the position of a $\delta$-ball inside which control is switched on)
can be stabilized by a difference control method that is 
similar to the LPLC method given above, even if one has only delayed
knowledge of differences of the system variables
$x_{t-\tau}-x_{t-\tau-1}$. Combined with stored values of the
meantime control amplitudes
$\vec{r}_1{\ldots}\vec{r}_{t-\tau-1}$ 
we propose the control scheme
\begin{eqnarray}
\vec{r}_t = K (\vec{x}_{t-\tau}-x_{t-\tau-1})  
 + \sum_{j=1}^{\tau+1} N_j \vec{r}_{t-j}
\end{eqnarray}
with the feedback matrices
\begin{eqnarray}
K &=& - M^{-1} L^{\tau+2} (L-\Einsop)^{-1} \\
\forall_{1\leq i \leq \tau} \;\;\;\;
N_i &=& - M^{-1} L^{i} M 
\\
N_{\tau+1} &=& M^{-1} L^{\tau+1} (L-\Einsop)^{-1} M
\end{eqnarray}
Here we have to assume that not only $M$ but also
% $(L-{1\hspace*{-0.35em}1})$ 
$(L-\Einsop)$
is invertible.
Using the linear approximation one easily computes directly that
this control leads to $\vec{x}_{t+1}=\vec{0}$, and again this is
a so-called deadbeat control scheme where all eigenvalues 
are zero.

Using the linearized dynamics, we have
\begin{eqnarray}
\vec{x}_{t-\tau} &=& L \vec{x}_{t-\tau-1} + M \vec{r}_{t-\tau-1}
\\ \vec{x}_{t} &=& L^{\tau+1}  \vec{x}_{t-\tau-1} 
+ \sum_{j=1}^{\tau+1} L^{j-1} M  \vec{r}_{t-j}
\end{eqnarray}
and the next iteration reads
\begin{eqnarray}
\vec{x}_{t+1} &=& L  \vec{x}_{t} + M  \vec{r}_{t}
=  L^{\tau+2} \vec{x}_{t-\tau-1}
+ \sum_{j=1}^{\tau+1} L^{j
} M  \vec{r}_{t-j}
\nonumber \\ &&
- M M^{-1}  L^{\tau+2} (L-\Einsop)^{-1} L \vec{x}_{t-\tau-1}
\nonumber \\ &&
- M M^{-1}  L^{\tau+2} (L-\Einsop)^{-1}  M \vec{r}_{t-\tau-1} 
\nonumber \\ &&
+ M M^{-1}  L^{\tau+2} (L-\Einsop)^{-1} \vec{x}_{t-\tau-1}
\nonumber \\ &&
- \sum_{j=1}^{\tau}  M M^{-1}  L^{j} M  \vec{r}_{t-j} 
\nonumber \\ &&
+ M M^{-1}   L^{\tau+1}  (L-\Einsop)^{-1} M \vec{r}_{t-\tau-1}
=\vec{0}
\end{eqnarray}
By this method the delay can be overcome and even fixed points
whose position is known inaccurate can be stabilized
from measured differences, provided that one has 
a sufficient number of control parameters, i.~e. 
$\dim\vec{r}=\dim\vec{x}$, and both $M$ and $(L-\Einsop)$
are invertible.

%\begin{references}

%\end{references}

% \onecolumn \include{changes}

\end{document}